\documentclass[12pt]{article}

\usepackage{latexsym}
\usepackage{amsfonts}
\usepackage{amssymb}
\usepackage{amsmath}

\usepackage{cite}

\font\tenmsbm=msbm10 scaled 1200
\font\sevenmsbm=msbm9
\newfam\msbmfam
\textfont\msbmfam=\tenmsbm \scriptfont\msbmfam=\sevenmsbm

\makeatletter
\@addtoreset{equation}{section}
\makeatother

\newcommand{\eref}[1]{(\ref{#1})}

\def\be{\begin{equation}}
\def\ee{\end{equation}}
\def\ba{\begin{eqnarray}}
\def\ea{\end{eqnarray}}
\def\bet{\begin{tabular}}
\def\eet{\end{tabular}}

\def\pa{\partial}

\def\ve{\varepsilon}

\def\vp{\varphi}
\def\ra{\rightarrow}

\def\a{\au}
\def\m{\mu}
\def\n{\nu}
\def\s{\sigma}
\def\a{\alpha}

\def\dl{\delta}

\def\wt{\widetilde}

\def\cM{{\cal M}}
\def\cL{{\cal L}}

\def\cN{{\cal N}}
\def\cE{{\cal E}}
\def\cB{{\cal B}}
\def\cK{{\cal K}}
\def\cV{{\cal V}}
\def\cU{{\cal U}}
\def\cS{{\cal S}}
\def\cF{{\cal F}}

\long\def\symbolfootnote[#1]#2{\begingroup
\def\thefootnote{\fnsymbol{footnote}}\footnote[#1]{#2}\endgroup}

\begin{document}

\begin{titlepage}
\null

\begin{center}

{\Large \bf Self-interacting chiral $p$-forms in higher dimensions}
\vskip2truecm

{\large Ginevra Buratti$^1$\symbolfootnote[1]{ginevra.buratti@uam.es},
Kurt Lechner$^{2,3}$\symbolfootnote[2]{kurt.lechner@pd.infn.it},
Luca Melotti$^2$\symbolfootnote[3]{luca.melotti.3@studenti.unipd.it}}

 \vspace{1.5cm}
$^1${\it Instituto de F\'isica Te\'orica UAM-CSIC, Cantoblanco, 28049 Madrid, Spain}

\vspace{1cm}
$^2${\it Dipartimento di Fisica e Astronomia ``Galileo Galilei''
\\
Universit\`a degli Studi di Padova,
Via Marzolo 8, 35131 Padova, Italy}

\vspace{1cm}
$^3${\it INFN, Sezione di Padova,
Via Marzolo 8, 35131 Padova, Italy}

\vspace{1.5cm}

\begin{abstract}

There exists no natural variational principle for the dynamics of abelian $p$-form potentials with self-dual field strengths, also called chiral $p$-forms. Relying on the PST method, we establish the general consistency condition for a Lagrangian to describe a Lorentz invariant self-interacting chiral $2n$-form in $4n+2$ dimensions. For a generic $n$, we determine a canonical solution of this condition for a quartic interaction Lagrangian of the $2n$-form, and prove that for the four-form in ten dimensions this interaction is unique. It generalizes the corresponding Born-Infeld-like interaction of a chiral two-form in six dimensions. We verify that under a dimensional reduction on a torus, this interaction Lagrangian reduces to a combination of the two recently constructed $SO(2)$-duality invariant quartic interactions for abelian three-form potentials in eight dimensions. The potential relevance of our method for the type IIB superstring effective action is discussed.

\end{abstract}

\end{center}

\vskip 1.0truecm \noindent {\it Keywords:} chiral p-forms, self-interactions, Lorentz invariance. {\it PACS:} 11.10.Kk, 11.10.Lm, 11.30.Rd, 11.30.Cp.
\end{titlepage}

\newpage

\baselineskip 6 mm


\tableofcontents

\section{Introduction}

The occurrence of self-dual gauge fields as possible elementary excitations of a consistent, classical or quantum, field theory has been hampered for a long time by the absence of a natural action describing their dynamics. The most prominent examples of fields of this kind are the chiral boson $\Lambda_0$ in two dimensions, the chiral two-form $\Lambda_2$ living on the $M5$-brane embedded in eleven-dimensional supergravity, and the chiral four-form $\Lambda_4$ belonging to the spectrum of the ten-dimensional type IIB superstring theory. The breakthrough occurred with the recognition that a non-manifestly Lorentz invariant action may produce a Lorentz invariant dynamics. This new strategy was pioneered by Floreanini and Jackiw \cite{Floreanini:1987as} for chiral bosons in two dimensions, and then extended to free chiral $2n$-forms $\Lambda$ in $4n+2$ dimensions by Henneaux and Teitelboim \cite{Henneaux:1988gg}. The topological properties of self-dual $2n$-forms in this non-manifestly invariant framework have been analyzed in \cite{Belov:2006jd}. A further step, within this approach, was the inclusion of self-interactions for chiral two-forms in a six-dimensional space-time, specifically the  Born-Infeld-type interaction on an $M5$-brane  \cite{Perry:1996mk, Schwarz:1997mc,Aganagic:1997zq}. More recently, this method has also been applied in \cite{Huang:2018hho} to the study of chiral two-forms interacting with non-abelian vector fields in six dimensions. In general, in this approach, the action is invariant under modified Lorentz transformations of the fields, which turn into the standard transformations on-shell, i.e. if one enforces the equations of motion.

Other approaches for self-dual gauge fields, instead of giving up manifest Lorentz invariance, rely on an action involving an infinite number of (typically massless) auxiliary fields, and try to reconcile their appearance with the standard paradigms of field theories with a finite number of degrees of freedom \cite{McClain:1990sx,Wotzasek:1990zr,Devecchi:1996cp,Bengtsson:1996fm,Berkovits:1996tn}.
In a specific case, an accurate choice of a single auxiliary tensor allowed to construct, via superspace techniques, a manifestly invariant equation of motion for the self-interacting chiral two-form of the $M5$-brane \cite{Howe:1996yn,Howe:1997fb} where, however, the reconstruction of the corresponding action is a rather non-trivial task \cite{Bandos:1997gm}. More recently, a manifestly Lorentz invariant approach has been introduced by Sen \cite{Sen:2015nph,Sen:2019qit}, anticipated in \cite{Saemann:2011nb,Mason:2011nw},
which relies on an action involving a single tensorial auxiliary field whose propagator has the wrong sign, but which eventually decouples from the physical degrees 
of freedom. Up to now, this approach has been applied to describe the interaction of chiral $2n$-forms with external fields, especially the dynamics of the four-form of type IIB supergravity, and to recover, via compactification, duality invariant versions of Maxwell theories in $4n$ dimensions. For a recent implementation of supersymmetry within this method, see \cite{Lambert:2019diy}.

Aim of this work is a general analysis of the allowed Lorentz-invariant self-interactions of chiral $2n$-forms in a space-time of generic dimension $D=4n+2$, including interactions with external fields and/or with charged sources. In general, the starting point cannot be a generic set of covariant (tensorial) equations of motion, since in general such a set does not represent a Poincar\'e invariant {\it fundamental} dynamics: as long as these equations do not follow from an {\it action}, they do, in fact, neither guarantee the basic conservation laws of four-momentum, angular momentum, and so on, nor do they ensure the existence of a Hamiltonian framework, necessarily preceding the quantization process. Hence, an efficient starting point is a variational principle based on an action. 

There exists a further approach for the construction of actions for chiral $2n$-forms, which is the Pasti-Sorokin-Tonin (PST) approach \cite{Pasti:1995tn,Pasti:1995us,Pasti:1996vs,Pasti:1997gx,Dall'Agata:1997ju,Bandos:1997ui,Dall'Agata:1997db,Dall'Agata:1998va,Lechner:1998ga}. It is naturally compatible with supersymmetry and $k$-symmetry, with the coupling to gravity, and with the functional integral approach. Its basic advantage is its manifest Lorentz invariance, realized via the introduction of a single scalar auxiliary field $a(x)$, which eventually must decouple via a shift-symmetry $a\ra a +\vp$, where $\vp$ is an arbitrary scalar field. The invariance of the action under this symmetry leads to a constraint for the Lagrangian, the so-called PST condition, which eventually guarantees the ``effective'' Lorentz invariance of the theory. For a recent variant of this method, see  \cite{Mkrtchyan:2019opf}.

In this paper, we derive the general PST condition \eref{PST} for a generic Lagrangian describing self-interacting chiral $2n$-forms. For definiteness, we consider only non-derivative couplings. For low space-time dimensions we retrieve as solutions of this condition the known dynamics for chiral bosons and two-forms in $D=2$ and $D=6$, respectively. For a generic space-time $D=4n+2$, we find a general class of solutions of the PST condition, that we call {\it canonical}, for Lorentz invariant quartic interaction Lagrangians for the field strength $H=d\Lambda$. For $D=10$, we prove that the canonical interaction is the {\it unique} quartic interaction solving the PST condition. Therefore, it represents the unique Lorentz invariant quartic interaction for a chiral four-form in $D=10$. However, unlike the six-dimensional case, the ten-dimensional quartic interaction is not of the Born-Infeld type, in the sense that it cannot be expressed in terms of the spatial components $B_{ijkl}\equiv H_{ijkl}$ of the field strength, and of its Hodge-dual $\wt B$, through the combinations, schematically, $(BB)^2$ and $(B\wt B)^2$.

Finally, we perform a non-trivial check of the consistency of this quartic interaction via a dimensional reduction on a torus from $D=10$ to $D=8$, which breaks the Lorentz group $SO(1,9)$ to $SO(1,7)\times SO(2)$. The resulting eight-dimensional theory is a manifestly $SO(2)$-duality invariant Maxwell theory for a couple of three-form potentials $A^I$, $I=1,2$, for which we have recently determined the {\it two} most general $SO(2)$-duality invariant quartic interactions \cite{Buratti:2019cbm}. We verify that the ten-dimensional quartic interaction Lagrangian of the chiral four-form goes indeed over in a sum of the two quartic interaction Lagrangians of the Maxwell field strengths $F^I=dA^I$. Via a Legendre transformation of the latter, we finally determine the corresponding quartic interaction of a single Maxwell field $F=dA$ in the manifestly 
Lorentz-invariant Gaillard-Zumino-Gibbons-Rasheed (GZGR) approach to duality \cite{Gaillard:1981rj,Gibbons:1995cv}, noting a remarkable coincidence. In the concluding Section \ref{outl}, we discuss the possible relevance of the new ten-dimensional quartic interaction, and of its generalizations, for the low energy effective action of type IIB superstrings, and we analyze potential extensions of our method to self-interactions of higher order.  

\section{Covariant action and PST consistency condition}
\label{caa}

We write the space-time dimension as $D=2p+2$, with $p$ {\it even}, and introduce the abelian $p$-form potential  $\Lambda_{\m_1\cdots\m_p}$. Its field strength is given by the antisymmetric tensor
\[
H_{\m_1\cdots\m_{p+1}}=(p+1)\,\pa_{[\m_1}\Lambda_{\m_2\cdots\m_{p+1}]},
\]
whose Hodge dual is defined in the usual manner as
\[
\wt H^{\m_1\cdots\m_{p+1}} =\frac{1}{(p+1)!}\,\ve^{\m_1\cdots\m_{p+1}\n_1\cdots\n_{p+1}}H_{\n_1\cdots\n_{p+1}},\quad\quad \wt{\wt H}=H.
\]
The equation of motion of a {\it free} (anti)chiral $p$-form is $\wt H=\pm H$. In the following, for definiteness, we will concentrate on chiral $p$-forms, $\wt H=H$. Chiral $p$-forms can be coupled to external fields via a {\it minimal} coupling, which amounts to impose the chirality condition on the modified field strength
\be
\label{Hfs}
H_{\m_1\cdots\m_{p+1}}=(p+1)\,\pa_{[\m_1}\Lambda_{\m_2\cdots\m_{p+1}]}+ \Phi_{\m_1\cdots\m_{p+1}},
\ee
where the antisymmetric tensor $\Phi_{\m_1\cdots\m_{p+1}}$ depends on the external fields, but not on the potential $\Lambda$ itself. For instance, in the $M5$-brane theory, $\Phi_{\m\n\rho}$ is the pull-back on the brane  worldvolume of the $D=11$ supergravity three-form potential $C_{MNK}$. In type IIB supergravity, $\Phi_{\m_1\cdots\m_5}$  is a combination of a Chern-Simons five-form, formed with the $D=10$ supergravity two-form potentials,
and of the gravitino and dilatino bilinears. If, on the other hand, the potential $\Lambda$ is coupled to a set of charged $(p-1)$-branes with charges $\{e_r\}$ -- for {\it chiral} $p$-forms electric and magnetic charges are identified -- then $\Phi$ is a linear combination of the $\delta$-functions $\Phi^r_{\m_1\cdots\m_{p+1}}$ supported on Dirac-$p$-branes, whose boundaries are the $(p-1)$-branes,
\be\label{diracp}
\Phi_{\m_1\cdots\m_{p+1}}=\sum_r e_r \Phi^r_{\m_1\cdots\m_{p+1}},\quad\quad J^r_{\m_1\cdots\m_p}=\pa^\m \wt \Phi^r_{\m\m_1\cdots\m_p}.
\ee
According to Poincar\'e duality, the current $J^r_{\m_1\cdots\m_p}$ is thus the $\dl$-function supported on the $r$-th $(p-1)$-brane. In this case, the field strength \eref{Hfs} satisfies the modified Bianchi identity $\pa_\m \wt H^{\m\m_1\cdots\m_p}=\sum_r e_r
J^{r\,\m_1\cdots\m_p}\equiv J^{\m_1\cdots\m_p}$.

In order to write a covariant action for a self-interacting chiral $p$-form, the PST approach foresees the introduction of a scalar auxiliary field $a(x)$ which allows, formally, to introduce a preferred direction of space-time $v^\m$, here assumed to be time-like,
\be
v^\m=\frac{\pa^\m a}{\sqrt{(\pa a)^2}},\quad \quad  v^2=1.
\ee
As anticipated, this scalar field must eventually become a pure-gauge degree of freedom.
Using the direction $v^\m$, we can introduce the ``electric" and ``magnetic" components of the field strength \eref{Hfs}
\be\label{ftch}
E_{\m_1\cdots\m_p}= H_{\m_1\cdots\m_p\m}v^\m,\quad\quad  B_{\m_1\cdots\m_p}=\wt H_{\m_1\cdots\m_p\m}v^\m,
\ee
which allow to decompose the field strength as 
 \be\label{Hdec}
H^{\m_1\cdots\m_{p+1}}=(p+1)\, E^{[\m_1\cdots\m_p}v^{\m_{p+1]}}+\frac{1}{p!}\,
\ve^{\m_1\cdots\m_{p+1}\n_1\cdots\n_{p+1}}B_{\n_1\cdots\n_p}v_{\n_{p+1}}.
\ee
The tensors $E$ and $B$ reduce to the actual electric and magnetic fields, if the auxiliary field is gauge-fixed to $a(x)=x^0$, see below, giving $v^\m=(1,0,\cdots,0)$.  Thanks to the relations \eref{ftch} and \eref{Hdec}, the equation for a free chiral field $H=\wt H$ is equivalent to the equality between the electric and magnetic fields, $E^{\m_1\cdots\m_p}=B^{\m_1\cdots\m_p}$. For a theory with non-linear interactions, we expect this equation to be modified to the non-linear first-order differential equation for the potential $\Lambda_{\m_1\cdots\m_p}$
\be\label{ebnonl}
E^{\m_1\cdots\m_p}=B^{\m_1\cdots\m_p}+g^{\m_1\cdots\m_p}(B),
\ee
where $g(B)$ is a local function of the magnetic field, which starts with cubic powers\footnote{\label{evenb}As equation \eref{ebnonl} must be Lorentz invariant, it should always be possible to put it in a manifestly Lorentz invariant form, say $H^{\m_1\cdots\m_{p+1}}=\wt H^{\m_1\cdots\m_{p+1}}+f_{\rm int}^{\m_1\cdots\m_{p+1}}(H)$, see e.g. \cite{Bandos:1997gm, Howe:1996yn}. Since $H$ and $f_{\rm int}$ are tensors of odd rank, $f_{\rm int}$ can contain only odd powers of $H$, and so the tensor $g(B)$ in \eref{ebnonl} can contain only {\it odd} powers of $B$, too.} of $B^{\m_1\cdots\m_p}$.
Understanding the contraction of the indices which are not written out explicitly, we have the further identities 
\be
HH=(p+1)(EE-BB)=-\wt H\wt H,\quad\quad H\wt H = 0.
\ee

We then propose as action for a self-interacting chiral $p$-form the functional of $\Lambda$ and $a$
\be\label{act}
I[\Lambda,a]=\frac{1}{p!}\int\left(\frac{1}{2}\,\big(EB +\Lambda J\big) -\cV(B)\right)d^Dx.
\ee
Here, for the moment, the {\it potential} $\cV(B)$ is a generic Lorentz invariant function of only the magnetic fields and $J$ is defined, for a generic function $\Phi$ of the external fields, by $J_{\m_1\cdots\m_p}=\pa^\m \wt \Phi_{\m\m_1\cdots\m_p}$. For the linear theory, the potential is given by $\cV(B)=\frac{1}{2}\,BB$, and in this case the action \eref{act} can be rewritten in the form 
\be\label{azlin}
I[\Lambda,a]=\frac{1}{2p!}\int\left( \frac{1}{2(p+1)}\,HH +\Lambda J -\frac{1}{2}\, hh\right)d^Dx, \quad \quad h^{\m_1\cdots\m_p}\equiv E^{\m_1\cdots\m_p}-B^{\m_1\cdots\m_p}.
\ee
We thus see that it differs from the action of a ``standard'' non-chiral field by a term which is proportional to the square of the self-duality relation $h^{\m_1\cdots\m_p}=0$. For the choice  $v^\m=(1,0,\cdots,0)$, the action \eref{azlin} reduces to the non-manifestly Lorentz-invariant actions for linear theories first given in \cite{Henneaux:1988gg,Deser:1997mz}.

To analyze the symmetries and equations of motion of the general action \eref{act} we write out its variation under generic variations $\dl\Lambda$ and $\dl a$
\be\label{vara}\begin{split}
\delta I[\Lambda,a]= -\frac{1}{(p!)^2}\int\ve^{\m_1\cdots\m_p}&^{\n_1\cdots\n_p \m\n}\bigg\{\pa_\n\big(h_{\m_1\cdots\m_p}v_\m\big)\dl \Lambda_{\n_1\cdots\n_p}\\[5pt]
&+
\bigg(\frac{1}{2}\,E_{\m_1\cdots\m_p}E_{\n_1\cdots\n_p}+\frac{1}{2}\,B_{\m_1\cdots\m_p} B_{\n_1\cdots\n_p}-V_{\m_1\cdots\m_p} E_{\n_1\cdots\n_p}\bigg)v_\m\dl v_\n\bigg\}d^Dx,
\end{split}
\ee
where we introduced the tensor 
\be\label{nb}
V_{\m_1\cdots\m_p}(B)= \frac{\pa \cV(B)}{\pa B^{\m_1\cdots\m_p}},
\ee
and $\dl v^\m= (\eta^{\m\n}-v^\m v^\n)\,\pa_\n\dl a/\sqrt{(\pa a)^2}$. From the above variation
one infers that, apart from the gauge symmetry $\dl\Lambda_{\m_1\cdots\m_p}=\pa_{[\m_1}\Sigma_{\m_2 \cdots\m_{p]}}$, the action is invariant under the {\it PST symmetries}
\be\label{pstsym}\begin{split}
\delta \Lambda_{\m_1\cdots\m_p}&=\frac{\vp}{\sqrt{(\pa a)^2}}\,
h_{\m_1\cdots\m_p} + \pa_{[\m_1}a\lambda_{\m_2\cdots\m_p]}, \\[5pt]
\delta a&=\vp,
\end{split}
\ee
where the local transformation parameters are the scalar field $\vp(x)$ and the tensor $\lambda_{\m_1\cdots\m_{p-1}}(x)$. Here we defined the ``generalized self-duality condition"
\be\label{heb}
h_{\m_1\cdots\m_p}=E_{\m_1\cdots\m_p}-V_{\m_1\cdots\m_p}(B).
\ee
The variation \eref{vara} vanishes trivially under the $\lambda$-symmetry of the gauge field $\Lambda_{\m_1\cdots\m_p}$, irrespective of the form of the potential $\cV(B)$. In particular, the magnetic field $B$ \eref{ftch} is inert under the $\lambda$-symmetry. Conversely, the $\vp$-symmetry, which shifts $a$ by an arbitrary scalar field, holds only for a subclass of potentials $\cV(B)$. In fact, inserting the transformations \eref{pstsym} in the variation \eref{vara} we find the compact expression (here one can replace $\dl v_\n\ra \pa_\n \vp/\sqrt{(\pa a)^2}$)
\be\label{varaz}
\delta I[\Lambda,a]= - \frac{1}{2(p!)^2}\int\ve^{\m_1\cdots\m_p\n_1\cdots\n_p\m\n}\left(B_{\m_1\cdots\m_p}B_{\n_1\cdots\n_p}-V_{\m_1\cdots\m_p}V_{\n_1\cdots\n_p}\right) v_\m\dl v_\n \, d^Dx.
\ee
Requiring that $\delta I[\Lambda,a]$ vanishes for generic $B$, $a$ and $\vp$, the validity of the 
$\vp$-symmetry imposes on $\cV(B)$ the {\it PST consistency condition} 
\be\label{PST}
\ve^{\m\n\m_1\cdots\m_p\n_1\cdots\n_p}\left(B_{\m_1\cdots\m_p}B_{\n_1\cdots\n_p}-V_{\m_1\cdots\m_p}V_{\n_1\cdots\n_p}\right) v_\m=0.
\ee
It can be written in the equivalent form $B_{[\m_1\cdots\m_p}B_{\n_1\cdots\n_p]}-V_{[\m_1\cdots\m_p}V_{\n_1\cdots\n_p]}=0$, since the tensors $B$ and $V$ are orthogonal to the vector $v^\m$. The PST condition \eref{PST} is thus the necessary and sufficient condition for $a(x)$ to be a pure-gauge degree of freedom, which can be fixed to an arbitrary (non-singular) value. In other words, once the field $a$ has been gauge fixed, the condition \eref{PST} ensures that the resulting theory is Lorentz invariant, despite this invariance is no longer manifest. We will discuss a class of relevant solutions of the condition \eref{PST} in Section \ref{solpst}. The linear theory, for which $\cV(B)=\frac{1}{2}\,BB$, satisfies this condition trivially.

\vskip0.3truecm
\noindent
{\it Equations of motion.} From \eref{vara} we read off the equations of motion for $\Lambda$ and $a$, respectively,
\begin{align}
\pa_{[\m_1}\left(v_{\m_2} h_{\m_3\cdots\m_{p+2}]}\right)&=0,\label{eqmai}\\[8pt]
\ve^{\rho\s\m_1\cdots\m_p\n_1\cdots\n_p}\,\pa_\rho\bigg(\frac{\pa_\s a}{(\pa a)^2}\,
h_{\m_1\cdots\m_p}\,h_{\n_1\cdots\n_p}\bigg)&=0,\label{eqma}
\end{align}
where in the equation of motion for $a$ we have used the PST condition \eref{PST}. By direct inspection, one sees that \eref{eqma} is a consequence of \eref{eqmai}. This was to be expected, since a pure-gauge field cannot imply any dynamics. We are thus left with the $\Lambda$-equation of motion \eref{eqmai}, whose general solution is $v_{[\m_1}h_{\m_2\cdots\m_{p+1}]}=\pa_{[\m_1}a\,\pa_{\m_2}\chi_{\m_3\cdots\m_{p+1]}}$, for some tensor $\chi$ of rank $p-1$. On the other hand, under the $\lambda$-transformation  \eref{pstsym} the left hand side of this relation transforms by $\dl(v_{[\m_1}h_{\m_2\cdots\m_{p+1}]})=-\pa_{[\m_1}a\,\pa_{\m_2}\lambda_{\m_3\cdots\m_{p+1]}}$. This implies that, by choosing $\lambda=\chi$, the equation of motion of the gauge field reduces to $v_{[\m_1}h_{\m_2\cdots\m_{p+1}]}=0 \leftrightarrow h_{\m_1\cdots\m_p}=0$. Within this gauge fixing, in light of \eref{heb} the equation of motion for $\Lambda$ corresponds  thus to the envisaged first-order generalized self-duality relation, see \eref{nb}, 
\be\label{eqchiral}
E_{\m_1\cdots\m_p}= V_{\m_1\cdots\m_p}(B).
\ee
Due to the normalization of the minimal-interaction term in the action \eref{act}, according to a classical argument \cite{Deser:1997mz}, see \cite{Lechner:2000eg} for a detailed derivation, a change of the Dirac-$p$-branes \eref{diracp} results in a change of the action by $I[\Lambda,a]\ra I[\Lambda,a]+ \frac{1}{2}\sum_{r,s}N_{rs}\,{e_re_s}$, where the $N_{rs}$ are integers. This implies that the charges of the $(p-1)$-branes must satisfy the quantization conditions
\be\label{diracq}
e_re_s=4\pi n_{rs},\quad n_{rs}\in \mathbb{Z}.
\ee

\section{Exact self-interactions in low dimensions}\label{solpst} 

\vskip0.3truecm
\noindent
{\it D=2.} In a two-dimensional space-time we have $p=0$, and a chiral $p$-form is a self-dual scalar field (or chiral boson) $\Lambda(x)$, whose field strength is the vector $H_\m=\pa_\m\Lambda$. Correspondingly, the electric and magnetic fields are also scalars, $E=v^\m\pa_\m\Lambda$, $B=v_\m\ve^{\m\n}\pa_\n\Lambda$. This case is in some sense special, as the $\lambda$-symmetry \eref{pstsym} assumes a slightly different form. To derive it, we apply a gauge transformation to rewrite it as $\dl\Lambda_{\m_1\cdots\m_p}= -a \,\pa_{[\m_1}\lambda_{\m_2\cdots\m_p]}$. Locally, in the language of differential forms, this transformation can be rephrased by saying that the variation of the $p$-form $\Lambda$ must be $a$ multiplied by a closed $p$-form. For a zero-form $\Lambda$, this implies that the $\lambda$-transformation takes the form $\dl\Lambda(x)=-\lambda a(x)$, where $\lambda$ is a constant. This time, the equation of motion \eref{eqmai} reads $\pa_{[\m}(v_{\n]}h)=0$, with general solution $v_\n h=\chi\pa_\n a$, for some constant $\chi$. Under a $\lambda$-symmetry we have $\dl(v_\n h)=-\lambda \pa_\n a$, so that by choosing $\lambda=\chi$ the generalized self-duality condition $h=E-V= E-d\cV(B)/dB$ can again be made to vanish.
However, the PST condition \eref{PST} in $D=2$ reads $\ve^{\m\n}(BB-VV)v_\m=0\Leftrightarrow 
B=d\cV(B)/dB$, which has as unique solution the linear theory $\cV(B)=\frac{1}{2}\,BB$, first analyzed in \cite{Floreanini:1987as}. Therefore, there exists no Lorentz-invariant self-interacting theory for a chiral boson in two dimensions.

\vskip0.3truecm
\noindent
{\it D=6.} The general self-interactions for a chiral two-form $\Lambda_{\m\n}$ in a six-dimensional space-time have first been analyzed in \cite{Perry:1996mk}. In this case, there exist two independent Lorentz-invariants that can be formed with the magnetic fields $B_{\m\n}$.
A convenient choice is 
\be\label{q12}
Q_1=\frac{1}{2}\,BB,\quad\quad Q_2= \frac{1}{16}\left(4{\rm tr}B^4-(BB)^2\right)=\frac{1}{16}\left(W^\m W_\m+(BB)^2\right).
\ee
In the last expression we have introduced the vector
\be\label{Wmu}
W^\m=-\frac{1}{2}\,\ve^{\m\n\m_1\m_2\n_1\n_2}B_{\m_1\m_2}B_{\n_1\n_2} v_\n, 
\ee
which appears also in the PST condition \eref{PST}. It entails the identities, specific for $D=6$,
\be\label{wprop}
W^\m W_\m=4{\rm tr}B^4-2(BB)^2, \quad\quad B^{\m\n}W_\m=0. 
\ee
Denoting the derivatives of the potential $\cV(Q_1,Q_2)$ by $V_i=\pa \cV(Q_1,Q_2)/\pa Q_i$, the PST condition \eref{PST} translates into
\be\label{PST6}
\frac{1}{2}\,\ve^{\m\n\m_1\m_2\n_1\n_2}\big(B_{\m_1\m_2}B_{\n_1\n_2}-V_{\m_1\m_2}V_{\n_1\n_2}\big)v_\m=\big(1-V_1^2+Q_2V_2^2\big)W^\n.
\ee
In this derivation, the second identity in \eref{wprop} is crucial to eliminate from the l.h.s. of \eref{PST6} a term which is not proportional to $W^\n$.
Lorentz invariance thus constrains the potential to satisfy the differential equation, first derived in \cite{Perry:1996mk} with a different choice of variables,
\be\label{perschw}
V_1^2-Q_2V_2^2=1.
\ee
This equation allows for an infinite set of solutions for $\cV(Q)$, see e.g. \cite{Perry:1996mk}. Particular examples are the free potential $\cV_0(Q)=Q_1$, and the Born-Infeld-like Lagrangian 
\be\label{bil}
\cV_{\rm BI}(Q)=\frac{2}{\gamma}\sqrt{\left(1+\frac{\gamma}{2}\,Q_1\right)^2-\gamma^2 Q_2}=\frac{2}{\gamma}\sqrt{\det\big(\delta^\m{}_\n+\sqrt{\gamma}B^\m{}_\n\big)}\,.
\ee
Up to terms of the eighth power in $B$, the general solution of equation \eref{perschw} has the universal expression 
\be
\cV(Q)=Q_1-\gamma Q_2+\frac{1}{2}\,\gamma^2Q_1Q_2+O(B^8).
\ee
In particular, the invariant $Q_2$ in \eref{q12} represents the unique quartic interaction in $D=6$.

\section{Quartic self-interactions in arbitrary dimensions}\label{quarticint} 

For dimensions $D=2p+2\ge 10$, a closed analytic way of writing the PST condition, analogous to \eref{PST6}, is no longer available. The invariants $Q_1$ and $W^\m W_\m$ in equations \eref{q12} and \eref{wprop}, respectively, can still be defined, where now
\be\label{Wd}
W^\m=-\frac{1}{p!}\,\ve^{\m\n\m_1\cdots\m_p\n_1\dots\n_p} B_{\m_1\cdots\m_p}B_{\n_1\cdots\n_p}v_\n.
\ee
However, since for $p\ge 4$ we have $B^{\m_1\cdots\m_p}W_{\m_p}\neq 0$, now the PST condition \eref{PST} develops also terms which are not proportional to $W^\n$, unlike what happens for $D=6$ in equation \eref{PST6}.

This means that, to derive solutions of the PST condition, we must resort to more general invariants of the fields $B$, in particular to a more general class of {\it quartic} invariants. We list some of them -- the simplest ones -- which will become relevant in the following:
\begin{align}
\cU_1&=(BB)^2,\label{u1}\\[5pt]
\cU_2&=(B^\m B^\n)(B_\m B_\n),\label{u2}\\[5pt]
\cU_3&=(B^{\m\n} B^{\rho\s})(B_{\m\n} B_{\rho\s}),\label{u3}\\[5pt]
\cU_4&=(B^{\m\n} B^{\rho\s})(B_{\m\rho} B_{\n\s})\label{u4}.
\end{align}
A further quartic invariant is given by the square of the vector \eref{Wd}
\be\label{ww}
W^\m W_\m= -\frac{(2p)!}{(p!)^2}\,B_{[\m_1\cdots\m_p}B_{\n_1\cdots\n_p]}B^{\m_1\cdots\m_p}B^{\n_1\cdots\n_p}.
\ee  
In \eref{u1}-\eref{u4} our convention is that the unwritten indices are contracted,
\be\label{convent}
(B^{\m\n} B^{\rho\sigma})=B^{\m\n\n_1\cdots\n_{p-2}}B^{\rho\sigma}
{}_{\n_1\cdots\n_{p-2}},\quad {\rm etc.}
\ee
Due to the growing complexity of the tensorial analysis in higher dimensions, we focus our attention to the quartic interactions. For this purpose, we expand the potential $\cV(B)$ in a power series in $B$ with the free part as the lowest order contribution
\be\label{uv}
\cV(B)=\frac{1}{2}\,BB+\cU(B)+O(B^6),
\ee
where $\cU(B)$ collects all quartic contributions (see Footnote \ref{evenb} in Section \ref{caa}). Plugging this expansion into the PST condition \eref{PST}, we find that $\cU(B)$ must satisfy the differential equation
\be
\ve^{\m\n\m_1\cdots\m_p\n_1\cdots\n_p}B_{\m_1\cdots\m_p}
\,\frac{\pa\,\cU(B)}{\pa B^{\n_1\cdots\n_p}}\,
v_\m=0. 
\ee
This equation can be recast as an invariance condition for the potential $\cU(B)$ if we introduce a {\it formal} transformation parameter $\Delta^\m$:
\be\label{deltau}
\delta\, \cU(B)=0, \quad\quad \delta B^{\m_1\cdots\mu_p}= \frac{1}{p!}\,\Delta_\n\, \ve^{\n\m\m_1\cdots\m_p\n_1\cdots\n_p}B_{\n_1\cdots\n_p}v_\m.
\ee
The search for solutions of the equation $\delta \,\cU(B)=0$ can be further simplified by choosing a Lorentz frame where $v^\m=\delta^{\m0}$. Then the magnetic field has only the spatial components $B^{m_1\cdots m_p}$, $m=(1,\cdots,D-1)$, and we can introduce its $(D-1)$-dimensional Hodge dual together with its inverse (henceforth, we will raise and lower the indices with the Euclidean metric $\delta^{mn}$)
\[
\wt B^{m_1\cdots m_{p+1}}= \frac{1}{p!}\,\ve^{m_1\cdots m_{p+1}n_1\cdots n_p} B^{n_1\cdots n_p},\quad\quad
B^{n_1\cdots n_p}= \frac{1}{(p+1)!}\,\ve^{n_1\cdots n_p m_1\cdots m_{p+1}} \wt B^{m_1\cdots m_{p+1}}.
\]
In this frame, the transformation law \eref{deltau} becomes 
\be\label{dlb}
\delta B^{m_1\cdots m_p}=\Delta^k \wt B^{k m_1\cdots m_p},\quad\quad \delta \wt B^{m_1\cdots m_{p+1}}=(p+1)\Delta^{[m_1}B^{m_2\cdots m_{p+1}]},
\ee
and the vector \eref{Wd} has only the spatial components, see the notation \eref{convent},
\be\label{wm}
W^m=(\wt B^m B).
\ee

\vskip0.3truecm
\noindent
{\it Canonical quartic invariants.}  We now proceed to the construction of a canonical quartic invariant $\cU_{\rm c}(B)$, satisfying the PST condition $\dl\,\cU_{\rm c}(B)=0$. We call it {\it canonical} because it is present for all dimensions $D\ge 6$. In particular, in $D=6$ and $D=10$ this invariant will turn out to be the {\it unique} quartic interaction. We begin by computing the variations of the invariants \eref{u1}, \eref{u2} and \eref{ww} under the transformation \eref{dlb} (in the frame where $v^\m=\delta^{\m0}$)
\begin{align}
\dl\,\cU_1&=4 \Delta^m (BB)W^m \label{du1}\\[5pt]
\dl\,\cU_2&=\frac{2}{p}\,\Delta^m\big((BB)W^m-(B^mB^n)W^n\big), \label{du2}\\[5pt]
\dl\big(W^\m W_\m\big)&=4\Delta^m\big(-(BB)W^m+p(B^mB^n)W^n\big). \label{dww}
\end{align}
The variation \eref{du1} follows from \eref{dlb} and \eref{wm}. The variation \eref{du2} relies on the identity
\[
(B^m\wt B^{kn})=\frac{1}{2p}\left(\dl^{mn}W^k-\dl^{mk}W^n\right),
\]
while the variation \eref{dww} is straightforward. What makes the invariants $\cU_1$, $\cU_2$ and $W^\m W_\m$ special is that their variation involves always the vector $W^m$ \eref{wm}, or $W^\m$ \eref{Wd}. From the above variations we see that the invariant which satisfies the PST condition $\dl\,\cU_{\rm c}(B)=0$ in any dimension is given by the combination
\be\label{uc}
\cU_{\rm c}(B)=(p-1)\,\cU_1-2p^2\,\cU_2-W^\m W_\m.
\ee
Notice that, unlike the six-dimensional universal quartic invariant $Q_2$ in \eref{q12}, for $D\ge 10$ Lorentz invariance requires in  $\cU_{\rm c}(B)$ the presence of the invariant $\cU_2$, in addition to $(BB)^2$ and $W^\m W_\m$. The dimension $D=6$ is special, because there we have the identity $(B^mB^n)W^n=0$ and, in addition, the invariant \eref{ww} is linearly dependent, $W^\m W_\m= 4\,\cU_2-2\,\cU_1$, see the relations \eref{wprop}. Nevertheless, also in this case the polynomial \eref{uc} reduces to the correct unique quartic invariant: $\cU_{\rm c}(B)=3\cU_1-12\cU_2=-48Q_2$. The dimension $D=2$ is even more special --  equation \eref{du2} must be replaced with $\dl\,\cU_2= 4 \Delta^m (BB)W^m $ -- but formula \eref{uc} still gives the expected result $\cU_{\rm c}(B)=0$.

\vskip0.3truecm
\noindent
{\it Uniqueness in $D=10$.} For a generic space-time dimension $D=4p+2$, the canonical quartic invariant $\cU_{\rm c}(B)$ is a rather complicated expression as the square $W^\m W_\m$ becomes the sum of a variety of quartic invariants. Moreover, for a generic $D\ge10$ {\it a priori} there is no reason to expect this quartic invariant to be unique. In both these respects, the dimension $D=10$ is special in that $i)$ $\cU_{\rm c}(B)$ has a simple form, and $ii)$ it is the {\it unique} quartic invariant solving the PST condition $\dl\, \cU(B)=0$. To see it, we write out the contractions appearing in \eref{ww}, for the case $p=4$, in terms of the polynomials \eref{u1}-\eref{u4}
\be
W^\m W_\m=-2\,\cU_1+32\,\cU_2-36 \,\cU_3.
\ee
The polynomial \eref{uc} then becomes
\be\label{ucten}
\cU_{\rm c}(B)=5\,\cU_1-64\,\cU_2+36 \,\cU_3.
\ee
The basic point is that the linearly independent quartic contractions of an antysimmetric four-tensor $B^{m_1m_2m_3m_4}$ are given by the {\it four} invariants \eref{u1}-\eref{u4}, but until now we did not consider $\cU_4$. To make a preliminary analysis, we assume that $B^{m_1m_2m_3m_4}$ has no components, say, along the ninth spatial direction, $m=9$. In this case, the equation $\dl \,\cU(B)=0$ becomes an eight-dimensional equation, where now $\dl B^{m_1m_2m_3m_4}=\Delta^9\wt B^{m_1m_2m_3m_4}$ is (proportional to) the eight-dimensional Hodge dual of the field. For this simpler configuration, the condition  $\dl\, \cU(B)=0$ is algebraically identical to the condition for a four-form field-strength $F^{\m\n\rho\sigma}$ in an eight-dimensional space-time to allow for a duality-invariant quartic interaction \`a la Gaillard-Zumino-Gibbons-Rasheed (GZGR), see \cite{Buratti:2019cbm}. Also for this last case, there are {\it a priori} four possible quartic polynomials, given precisely by \eref{u1}-\eref{u4} with the replacement $B^{m_1m_2m_3m_4}\ra F^{\m\n\rho\sigma}$. But for that case in \cite{Buratti:2019cbm} it has been shown that $\cU_4(F)$, or any of its combinations with $\cU_1(F)$, $\cU_2(F)$ and $\cU_3(F)$, cannot satisfy the condition $\dl \,\cU(F)=0$. This implies, a fortiori, that a combination involving $\cU_4(B)$ cannot satisfy the (more stringent) PST condition \eref{deltau}. It follows that the polynomial \eref{ucten} is the unique possible quartic interaction for a chiral four-form in a ten-dimensional space-time.

\section{Dimensional reduction on a torus}

From a kinematical point of view, via a dimensional reduction on a torus of equal periods, a theory of chiral $p$-forms in $2p+2$ dimensions goes over to an $SO(2)$-duality invariant theory of a pair of Maxwell-like $(p-1)$-form potentials $A^I_{\m_1\cdots\m_{p-1}}$ in $D=2p$ dimensions ($I=1,2$). The latter type of theories, in all examples that have been explicitly worked out for $D=4$ in the literature, see for instance \cite{Deser:1997gq,Buratti:2019cbm,Bossard:2011ij,Carrasco:2011jv,Pasti:2012wv}, turn out to be equivalent, in turn, to a duality-invariant theory formulated \`a la GZGR \cite{Gaillard:1981rj,Gibbons:1995cv} in terms of a {\it single} potential $A_{\m_1\cdots\m_{p-1}}$. Recently, we have provided the first, canonical and non-canonical, quartic interactions for a duality invariant pair of three-form potentials $A^I_{\m_1\m_2\m_3}$ in $D=8$, and we have proven their equivalence with a corresponding new class of quartic interactions in the GZGR formulation \cite{Buratti:2019cbm}. In this section, we will establish the relation between the ``chiral" quartic interaction \eref{ucten} in $D=10$ and these $SO(2)$-duality-invariant interactions in $D=8$. 

Before proceeding, we briefly recall the general construction of a manifestly $SO(2)$-invariant theory of a pair of self-interacting $(p-1)$-form potentials $A^I$ in $D=2p$ \cite{Buratti:2019cbm}, based again on the PST method. We introduce the 
pair of field strengths 
\be\label{fim}
F^I_{\m_1\cdots\m_p}=p \,\pa_{[\m_1}A^I_{\m_2\cdots\m_p]}+ \Phi^I_{\m_1\cdots\m_p},
\ee 
and the corresponding pairs of electric and magnetic fields ${\cE}^I_{\m_1\cdots\m_{p-1}}= F^I_{\m_1\cdots\m_p}v^{\m_p}$ and ${\cB}^I_{\m_1\cdots\m_{p-1}}=\wt F^I_{\m_1\cdots\m_p}v^{\m_p}$. The action then takes the form
\be
\label{saa}
S[A,a]=\frac{1}{(p-1)!}\int\left(\frac{1}{2}\,\varepsilon^{IJ}\left(\cE^I \cB^J+A^IJ^J\right) +\cN(\cB)\right)d^{2p}x,
\ee
where the unwritten indices are contracted, as in the action \eref{act}, and the {\it Hamiltonian} $\cN(\cB)$ is a Lorentz and $SO(2)$-invariant function of only the magnetic tensors ${\cB}^I_{\m_1\cdots\m_p}$. If the potentials $A^I$ are coupled to a set of dyonic $(p-2)$-branes
with electric and magnetic charges $\{e^1_r,e^2_r\}$, the current-doublet in \eref{saa} has the expression $J^I_{\m_1\cdots\m_{p-1}}=\sum_r e_r^I\,J^r_{\m_1\cdots\m_{p-1}}=
\sum_r e_r^I \, \pa^\m  \wt\Phi^r_{\m\m_1\cdots\m_{p-1}}$, see  \eref{diracp}, and then the doublet representing the sources in \eref{fim} is given by $\Phi^I_{\m_1\cdots\m_p}=\sum_r e_r^I\,\Phi^r_{\m_1\cdots\m_p}$. Again, a standard argument \cite{Deser:1997mz} shows that under a change of the $\dl$-functions $\Phi^r_{\m_1\cdots\m_p}$, i.e. a change of the Dirac-$(p-1)$-branes, the action \eref{saa} changes by an integer multiple of $2\pi$, if the charges satisfy Schwinger's quantization conditions
\be\label{schw}
\left(\,e_r^2e_s^1-e_r^1e_s^2\,\right)=4\pi n_{rs}, \quad n_{rs}\in {\mathbb Z}.
\ee
This time, the invariance of the action $S[A,a]$ under the PST symmetries requires $\cN(\cB)$ to satisfy the differential equation 
\be\label{PSTdual}
\ve^{\m\n\m_1\cdots\m_{p-1}\n_1\cdots\n_{p-1}}\,\ve^{IJ}\left(
\cB^I_{\m_1\cdots\m_{p-1}}\,\cB ^J_{\n_1\cdots\n_{p-1}}- N^I_{\m_1\cdots\m_{p-1}}\,
N^J_{\n_1\cdots\n_{p-1}}\right)v_\m=0,
\ee
where $N^I_{\m_1\cdots\m_{p-1}}(\cB)=\pa \cN(\cB)/\pa\cB^{I\m_1\cdots\m_{p-1}}$. The gauge-fixed equation of motion, which expresses the electric fields in terms of the magnetic ones, is given by $\cE^I_{\m_1\cdots\m_{p-1}}=-\ve^{IJ} N^J_{\m_1\cdots\m_{p-1}}(\cB)$, to be compared with the generalized self-duality equation \eref{eqchiral}. Key point of the approach is again the solution of the PST condition \eref{PSTdual}, which ensures the decoupling of $a(x)$, i.e. Lorentz invariance. If we expand the Hamiltonian in a power series in $\cB$, with the free part as the lowest contribution,
\be\label{nm}
\cN(\cB)=\frac{1}{2}\,\cB^I\cB^I+\cM(\cB)+O(\cB^6),
\ee
as in \eref{uv}, then in \cite{Buratti:2019cbm} it has been shown that for $D=8$ the most general $SO(2)$-invariant quartic interaction $\cM(\cB)$ for the magnetic field $\cB^I_{\m_1\m_2\m_3}$ -- solution of the PST condition \eref{PSTdual} -- is a linear combination of two independent polynomials
\be\label{ms12}
\cM(\cB)=c_1\,S_1(\cB)+c_2\,S_2(\cB),
\ee
given by, see the notation \eref{convent},
\begin{align}
S_1(\cB)&= \left(\cB^I \cB^J\right)(\cB^I \cB^J)-\frac{1}{2}\left(\cB^I \cB^I\right)^2,\\[5pt]
S_2(\cB)&=\left(\cB^I \cB^I\right)^2-12\left(\cB^{I\m} \cB^{I\n}\right)\left(\cB^J{}_\m \cB^J{}_\n\right)+6 \left(\cB^{I\m} \cB^{J\n}\right)\left(\cB^I{}_\m \cB^J{}_\n\right).
\end{align}

\subsection{From chiral $p$-forms to duality-invariant $(p-1)$-forms.} 

We now move to the reduction of the action \eref{act} for a chiral $p$-form on a torus. To preserve $SO(2)$-invariance, we take the two handles of the torus to have the same length $L=1$. Henceforth, we denote the $(2p+2)$-dimensional indices with a  bar, $\bar\m=(\mu,I)$, where $\mu=(0,\cdots,2p-1)$, and $I=(2p,2p+1)$ denotes the two space-like compact directions. The coordinates $x^I$ vary thus between $0$ and $1$. We consider as the only non-vanishing components of the potential $\Lambda_{\bar\m_1\cdots\bar\m_p}(\bar x)$ the mixed components $\Lambda_{I\m_1\cdots\m_{p-1}}(x)$, that we take to depend only on the non-compact directions $x^\m$. These components identify directly the pair of $(p-1)$-form potentials in $D=2p$ as $A^I_{\m_1\cdots\m_{p-1}}=-\Lambda_{I\m_1\cdots\m_{p-1}}$. The unique non-vanishing components of the field strength \eref{fim} are then given by\footnote{The reduction of the Dirac-brane $\delta$-functions $\Phi_{\m_1\cdots\m_{p+1}}$ can be implemented as exemplified in reference \cite{Deser:1997mz}.}
\be\label{hf}
H_{I\m_1\cdots\m_p}=-p\,\pa_{[\m_1}\Lambda_{I\m_2\cdots\m_p]}+\Phi_{I\m_1\cdots\m_p}\equiv F^I_{\m_1
\cdots\m_p},
\ee
which are thus identified with the $2p$-dimensional field strengths \eref{fim}. Correspondingly, the electric and magnetic fields \eref{ftch} of the $(2p+2)$-dimensional field strength $H$ reduce to the electric and magnetic components of the field strengths $F^I$
\be\label{ebred}
E_{I\m_1\cdots\m_{p-1}}=\cE^I_{\m_1\cdots\m_{p-1}},\quad\quad B^{I\m_1\cdots\m_{p-1}}=\ve^{IJ}\cB^{J\m_1\cdots\m_{p-1}}.
\ee
In this way, the first term of the action \eref{act} becomes 
\be
EB=p\,\varepsilon^{IJ}\cE^I \cB^J.
\ee
The $(2p+2)$-dimensional action \eref{act} then indeed reduces to a $2p$-dimensional action of the form \eref{saa}
\be\label{ired}
I[\Lambda,a]\quad\rightarrow\quad S[A,a],
\ee
where the Hamiltionian appearing in the latter is given by
\be\label{nv}
\cN(\cB)=-\frac{1}{p}\,\cV\big(\ve \cB\big).
\ee
For the reduction of the minimal-interaction term, see the end of the section. As $\cV(B)$ is Lorentz invariant in $2p+2$ dimensions, $\cN(\cB)$ is Lorentz invariant in $2p$ dimensions as well as $SO(2)$-duality-invariant. Moreover, with the reductions \eref{ebred}, the PST condition \eref{PST} for $\cV\big(B)$ goes over precisely to the PST condition \eref{PSTdual} for the Hamiltonian $\cN(\cB)$ \eref{nv}: in other words, the decoupling of $a(\bar x)$ in $2p+2$ dimensions implies the decoupling of $a(x)$ in $2p$ dimensions.

Comparing equations \eref{uv}, \eref{nm} and \eref{nv}, we see that the quartic deformation of the $2p$-dimensional theory is given in terms of the quartic deformation of the $(2p+2)$-dimensional theory by
\be\label{mu}
\cM(\cB)=-\frac{1}{p}\,\cU\big(\ve \cB\big).
\ee
The correspondence between the PST conditions in $D=2p+2$ and $D=2p$ then implies that the canonical quartic interaction $\cU_{\rm c}(B)$ \eref{ucten} for chiral four-forms in $D=10$, once reduced from $D=10$ to $D=8$ according to \eref{ebred}, satisfies automatically the PST condition \eref{PSTdual} for duality-invariant $(p-1)$-forms. As the general solution of the latter is given by the combination \eref{ms12}, the reduced quartic interaction $\cU_{\rm c}\big(\ve \cB\big)$ must necessarily be a combination of this type. We can make an explicit check of this general property, by replacing in \eref{ucten} the field $B^{\bar\m_1\bar\m_2\bar\m_3\bar\m_4}$ with $B^{I\m_1\m_2\m_3} =\ve^{IJ}\cB^{J\m_1\m_2\m_3}$. An explicit calculation yields indeed the expected result
\be
\cM_{\rm c}(\cB)=-\frac{1}{4}\,\cU_{\rm c}\big(\ve\cB\big)=16\,S_1(\cB)-12\, S_2(\cB),
\ee
i.e. $c_1=16$ and $c_2=-12$. 

We can make a further step forward by recalling that, as shown in \cite{Buratti:2019cbm}, the manifestly duality invariant quartic interaction \eref{ms12} can be mapped in a manifestly Lorentz invariant GZGR-type Lagrangian $\cL(F)$ of a single field strength $F_{\m_1\m_2\m_3\m_4}=4\pa_{[\m_1}A_{\m_2\m_3\m_4]}+ \Phi_{\m_1\m_2\m_3\m_4}$ in $D=8$, for which duality is only a symmetry of the equations of motion. The Lagrangian corresponding to \eref{ms12} turned out to be given by $\cL(F)=-\frac{1}{2\cdot 4!}\,FF+\cK(F)$, where
\be\label{k12}
\cK(F)=\frac{1}{(4!)^2}\left(3c_1+2c_2\right)R_1(F)-\frac{c_2}{9}\,R_2(F),
\ee
and $R_1(F)$ and $R_2(F)$ are the duality-invariant quartic polynomials
\begin{align}
R_1(F)&= \big(F\wt F\big)^2+\big( FF\big)^2,\label{r1}\\[5pt]
R_2(F)&= \big(F^\m F^\n\big)\big(F_\m F_\n\big)-\frac{1}{8}\,\big(FF\big)^2.
\end{align}
Writing out the Hodge dual in \eref{r1}, with the above values of $c_1$ and $c_2$ the 
quartic interaction \eref{k12} becomes
\be\label{kgz}
\cK(F)= -\frac{5}{24}\,\big(FF\big)^2 +\frac{8}{3}\,\big(F^\m F^\n\big)\big(F_\m F_\n\big)-\frac{3}{2}\,\big(F^{\m\n}F^{\rho\sigma}\big)\big(F_{\m\n}F_{\rho\sigma}\big).
\ee
In conclusion, the Lorentz invariant quartic interaction \eref{ucten} of a chiral four-form $\Lambda_{\m_1\m_2\m_3\m_4}$ in $D=10$ -- via a dimensional reduction on a torus, followed by an inverse Legendre transformation leading from the PST to the GZGR formulation -- gives rise to the particular duality-invariant quartic interaction \eref{kgz} for an abelian three-form potential $A_{\m_1\m_2\m_3}$ in $D=8$. Finally, by direct inspection of formulas \eref{u1}-\eref{u3}, \eref{ucten} and \eref{kgz}, we notice the, in a sense intriguing, {\it formal} coincidence
\be\label{kuc}
\cK(F)=-\frac{1}{4!}\,\cU_{\rm c}(F).
\ee
Until now we do not have any explanation for this remarkably simple relation between the quartic interactions of chiral four-forms in $D=10$ and duality invariant three-forms in $D=8$, nor do we know if it should be expected to hold in general for the reduction from $D=2p+2$ to $D=2p$.

\vskip0.3truecm
\noindent
{\it Reduction of the charged sources.}  For a consistent reduction of the charged sources the two compact coordinates of each $(p-1)$-brane must wrap a certain number of times the torus. We write the coordinates of the $r$-th brane as $x_r^{\bar\mu}(\sigma,\lambda)$, where the parameters $\sigma=(\sigma_1,\cdots,\sigma_{p-1})$ describe the world-volume of the reduced (dyonic) $(p-2)$-brane, and $\lambda\in[0,2\pi]$ is a compact parameter. We choose the $2p$-dimensional coordinates to be independent of $\lambda$, $x_r^\mu(\sigma,\lambda)= x_r^\mu(\sigma)$, while the compact ones $x^I_r(\sigma,\lambda)$ must satisfy the wrapping conditions
$\int_0^{2\pi}\frac{\pa x^I_r(\sigma,\lambda)}{\pa\lambda}\,d\lambda=N_r^I$, where the $N_r^I$ are integer winding numbers. Then the minimal-interaction term of the action \eref{act} can be seen to  reduce to $\int\Lambda J\,d^{2p+2}x=p\int \ve^{IJ}A^IJ^Jd^{2p}x$, where the $2p$-dimensional currents $J^I=\sum_r e_r^IJ_r$ carry the dyon charges $e_r^I=\ve^{IJ}N_r^Je_r$, earned from the windings. In this way, the action \eref{act} reduces thus as in \eref{ired}. As last consistency check we observe that, if the charges $e_r$ of the chiral $(p-1)$-branes satisfy the quantization conditions \eref{diracq}, then the charges $e_r^I=\ve^{IJ}N_r^Je_r$ of the dyonic 
$(p-2)$-branes automatically satisfy Schwinger's quantization conditions \eref{schw}.

\section{Outlook} \label{outl}

We have established the general condition \eref{PST} for the potential $\cV(B)$ of a self-interacting chiral $p$-form to represent a Lorentz invariant dynamics. In low dimensions, it reproduces the known theories, and for a generic dimension $D\ge 10$ we have established the canonical Lorentz invariant quartic interaction $\cU_{\rm c}(B)$, see \eref{uc}. In $D=10$, a double dimensional reduction of the latter gives rise to a specific $SO(2)$-duality invariant quartic interaction for a pair of three-form potentials in $D=8$, which agrees with a previous, independent, construction of these interactions. In the GZGR formulation, this interaction corresponds, unexpectedly, to the same formal quartic polynomial of the chiral four-form from which it originates, formula \eref{kuc}.

There are several issues left open. The first regards the question of higher-order interactions, say polynomials $\cS(B)$ of the sixth power in the fields, which deform the potential \eref{uv} into $\cV(B)=\frac{1}{2}\,BB+\cU(B)+\cS(B)+O(B^8)$. Supposing that the quartic interactions obey $\dl\,\cU(B)=0$, the PST condition \eref{PST} becomes a differential equation for the sixth-order interactions $\cS(B)$
\be\label{PSTz}
\dl\cS(B)=\frac{1}{2p!}\,\Delta_\n\,\ve^{\m\n\m_1\cdots\m_p\n_1\cdots\n_p}\,\frac{\pa\,\cU(B)}{\pa B^{\m_1\cdots\m_p}}\,\frac{\pa\,\cU(B)}{\pa B^{\n_1\cdots\n_p}}\,v_\m.
\ee
This means that each quartic polynomial $\cU(B)$ requires necessarily the presence of a non-vanishing sixth-order polynomial $\cS_0(B)$, satisfying equation \eref{PSTz}, to which one must add the most general solution of the homogeneous equation $\dl\cS_{\rm hom}(B)=0$, which may entail additional sixth-order polynomials.

One of the most interesting self-dual fields is the chiral four-form $\Lambda_{\m\n\rho\sigma}$ of type IIB superstring theory. For dimensional reasons, its low energy effective action, as an expansion in the string slope $\a'$, schematically gives rise to a potential, now depending also on a set of {\it external} fields $\vp=(R,\cF,\cdots)$, of the kind $\cV(B,\vp)= \frac{1}{2}\,BB+\a'B^4+ \a'^2 (\pa B)^2B^2+\a'^3\big((\pa B)^4+ B^4(\pa B)^2+ B^4R^2+
B^4 \cF^4+\cdots\big)$, where $R$ is the Riemann tensor and $\cF$ represents the doublet of the three-form field strengths of type IIB supergravity\footnote{In its {\it original} form $\cV(H,\vp)$, this effective action is usually reconstructed from on-shell superstring amplitudes, or deduced via on-shell supersymmetry arguments, both approaches where $H$ is intrinsically self-dual, $E=B$, see equations \eref{ftch}, \eref{Hdec}. The potential $\cV(B,\vp)$ is then derived from $\cV(H,\vp)$ by replacing again $E$ with $B$, eliminating thus the ambiguities which are intrinsic to $\cV(H,\vp)$.}. The dots stand for additional terms with the correct dimensions, involving also the scalars of the theory. In these effective interactions, the self-duality of the five-form field strength $H=d\Lambda+\Phi$ \eref{Hfs} is usually imposed on the equations of motion \cite{Policastro:2006vt}, or enforced by replacing in the effective action $H$ with its self-dual part, $H\ra\frac{1}{2}(H+\wt H)$ \cite{Peeters:2003pv}, thus renouncing to the control of Lorentz invariance. According to the explicit form of the low energy effective action, as far as known, see for instance \cite{Green:1998by,deHaro:2002vk,Peeters:2003pv,Green:2003an,Policastro:2006vt}, actually there are no order-$\a'$ and order-$\a'^2$ corrections up to one loop order in string theory, and they are conjectured to be absent at all orders. Conversely, would there be higher loop corrections of order $\a'$ involving the chiral four-form, then our analysis shows that they must necessarily be proportional to the polynomial $\cU_{\rm c}(B)$ \eref{ucten}. On the other hand, the polynomial $\cU_{\rm c}(B)$ must appear also in the order-$\a'^3$ corrections above, whenever the indices of the fields $B_{\m_1\m_2\m_3\m_4}$ are all contracted among four of them, i.e. if in $\cV(B,\vp)$ an interaction appears in the factorized form $\cU_f(B,\vp)=\a'^3 (B^4) f(\vp)$, where $f(\vp)$ is a scalar function of the other fields. In fact, in this case the condition \eref{deltau}, applied to $\cU_f(B,\vp)$, forces the prefactor $(B^4)$ again to be proportional to $\cU_{\rm c}(B)$. 

More generally, the above expansion of the potential $\cV(B,\vp)$, apart from involving the external fields, exhibits the new feature that $B$ appears now also via its derivatives. However, it is not difficult to see that for this more general case the Lorentz-invariance condition \eref{PST} keeps the same form: it suffices to replace the derivatives $V_{\m_1\cdots\m_p}(B)$ \eref{nb} with the functional derivatives $V_{\m_1\cdots\m_p}(B,\vp)={\dl \int\cV(B,\vp)\,d^{10}x}/\dl B^{\m_1\cdots\m_p}$.
Concrete applications of our method to the effective action of type IIB superstring theory requires, first of all, to develop a systematic technique to impose the condition \eref{PST} on  the potential $\cV(B,\vp)$, a challenge for future work.

Another issue left for future investigation is to write the equation of motion \eref{eqchiral}, following from the action \eref{act}, explicitly in a manifestly covariant form. In fact, using the decomposition \eref{Hdec}, separating the free and interaction parts one can always rewrite equation \eref{eqchiral} in the form $H-\wt H= f_{\rm int}(H,a)$, an equation between anti-self-dual tensors of rank $p+1$, see Footnote \ref{evenb} of Section \ref{caa}. However, the r.h.s. of the latter cannot depend on $a$, because with the gauge-fixing leading to \eref{eqchiral}, the PST symmetries \eref{pstsym} reduce to $\dl a = \vp$, $\dl \Lambda =0$, giving $\dl H=0$, see \eref{heb}. We thus obtain the manifestly invariant form $H-\wt H= f_{\rm int}(H)$. An interesting issue would be to determine the (cubic) tensor $f_{\rm int}(H)$ associated with the canonical quartic interaction $\cU_{\rm c}(B)$ \eref{uc}.

A further interesting question regards the extension of Sen's approach for chiral $p$-forms \cite{Sen:2015nph,Sen:2019qit}, and of the ``extended" PST approach developed recently by Mkrtchyan \cite{Mkrtchyan:2019opf}, to include self-interactions of the field. So far, these covariant approaches, which both require the introduction of an additional auxiliary (chiral) $p$-form, contemplate only actions involving at most quadratic terms of the field strength $H$.  

\vskip0.5truecm
\paragraph{Acknowledgments.}
This work is supported in part by the INFN CSN4 Special Initiative {\it STEFI}. K.L. thanks Dmitri Sorokin for useful discussions.


\providecommand{\href}[2]{#2}\begingroup\raggedright\endgroup
\end{document}